\documentclass[12pt, draftclsnofoot, onecolumn]{IEEEtran}

\usepackage[dvipsnames]{xcolor}
\usepackage{algpseudocode}
\usepackage{svg}
\usepackage{setspace}
\usepackage{amsthm}
\newtheorem{lemma}{Lemma}
\usepackage{amsmath}%
\usepackage{MnSymbol}%
\usepackage{wasysym}
\DeclareMathOperator*{\argmax}{arg\,max}

\usepackage{subcaption}
\usepackage{diagbox}

\DeclareMathAlphabet{\pazocal}{OMS}{zplm}{m}{n}

\usepackage[linesnumbered,ruled,vlined]{algorithm2e}

\usepackage[mathscr]{euscript}
\usepackage{cite}
\usepackage{psfrag}

\usepackage{soul}
\usepackage{amsmath}
\usepackage{array}
\usepackage{pifont}
\newcolumntype{P}[1]{>{\centering\arraybackslash}p{#1}}
\usepackage{tikz}
\usepackage[utf8]{inputenc}
\usepackage{pgfplots} 
\usepackage{pgfgantt}
\usepackage{pdflscape}
\pgfplotsset{compat=newest} 
\pgfplotsset{plot coordinates/math parser=false}
\usepackage{grffile}
\usetikzlibrary{plotmarks}
\usepackage{amsthm}
\usepackage{amsmath}%
\usepackage{MnSymbol}%
\usepackage{wasysym}
\usepackage{subcaption}
\usepackage[mathscr]{euscript}
\usepackage{cite}
\usepackage{amsfonts}
\usepackage{pgf}
\usepackage{mathtools}
\usepackage{graphicx}
\usepackage{array}
\usepackage{tikz}
\usepackage[utf8]{inputenc}
\usepackage{pgfplots} 
\usepackage{pgfgantt}
\usepackage{pdflscape}
\pgfplotsset{compat=newest} 
\pgfplotsset{plot coordinates/math parser=false}
\usepackage{grffile}
\usetikzlibrary{plotmarks}
\usepackage{tikzscale}
\usepackage{caption}
\usepackage{subcaption}
\usepackage{amsmath}
\usepackage{array}
\usepackage{tikz}
\usepackage[utf8]{inputenc}
\usepackage{pgfplots} 
\usepackage{pgfgantt}
\usepackage{pdflscape}
\pgfplotsset{compat=newest} 
\pgfplotsset{plot coordinates/math parser=false}
\usepackage{grffile}
\usetikzlibrary{plotmarks}
\usepackage[T1]{fontenc}
\usepackage{graphicx}
\usepackage{flushend}
\usepackage{caption}
\usepackage{overpic}
\usepackage{array}
\usepackage{color}
\usepackage{url}
\usepackage{tikz}
\usepackage[utf8]{inputenc}
\usepackage{pgfplots} 
\usepackage{cite}

\newtheorem{theorem}{Theorem}

\newcommand{\bigtriangleq}{\mathbin{\setstackgap{S}{0pt}\stackMath\Shortstack{\smalltriangleup\\ =}}}
\usepackage[usestackEOL]{stackengine}

\def\CN{\mathcal{C}\mathcal{N}} 

\usepackage[colorlinks=true,citecolor=blue,urlcolor=blue]{hyperref}  
\IEEEoverridecommandlockouts
\makeatletter
\def\endthebibliography{%
	\def\@noitemerr{\@latex@warning{Empty `thebibliography' environment}}%
	\endlist
}
\makeatother

\begin{document}
	\title{Deep Reinforcement Learning for Practical Phase Shift Optimization in RIS-aided MISO URLLC Systems}
	
	\author{Ramin~Hashemi,~\IEEEmembership{Student Member,~IEEE,}
	Samad Ali, \IEEEmembership{Member,~IEEE},
	Nurul Huda Mahmood, 
	  and Matti Latva-aho, \IEEEmembership{Senior Member,~IEEE}

	\thanks{The authors are with the Centre for Wireless Communications (CWC), University of Oulu, 90014 Oulu, Finland. e-mails: (\{ramin.hashemi,  samad.ali, nurulhuda.mahmood,  matti.latva-aho\}@oulu.fi).
	
	This research has been supported by the Academy of Finland, 6G Flagship program under Grant 346208.
	
		}
		}
	
	\maketitle

	\begin{abstract}
		We study the joint active/passive beamforming and channel blocklength (CBL) allocation in a non-ideal reconfigurable intelligent surface (RIS)-aided ultra-reliable and low-latency communication (URLLC) system. The considered scenario is a finite blocklength (FBL) regime and the problem is solved by leveraging a novel deep reinforcement learning (DRL) algorithm named twin-delayed deep deterministic policy gradient (TD3). First, assuming an industrial automation system with multiple actuators, the signal-to-interference-plus-noise ratio and achievable rate in the FBL regime are identified for each actuator in terms of the phase shift configuration matrix at the RIS. Next, the joint active/passive beamforming and CBL optimization problem is formulated where the objective is to maximize the total achievable FBL rate in all actuators, subject to non-linear amplitude response at the RIS elements, BS transmit power budget, and total available CBL. Since the amplitude response equality constraint is highly non-convex and non-linear, we resort to employing an actor-critic policy gradient DRL algorithm based on TD3. The considered method relies on interacting RIS with the industrial automation environment by taking actions which are the phase shifts at the RIS elements, CBL variables, and BS beamforming to maximize the expected observed reward, i.e., the total FBL rate. We assess the performance loss of the system when the RIS is non-ideal, i.e., with non-linear amplitude response, and compare it with ideal RIS without impairments. The numerical results show that optimizing the RIS phase shifts, BS beamforming, and CBL variables via the proposed TD3 method is highly beneficial to improving the network total FBL rate as the proposed method with deterministic policy outperforms conventional methods.
	\end{abstract}
	
	
	\IEEEpeerreviewmaketitle
	
	\section{Introduction}
	\bstctlcite{IEEEexample:BSTcontrol}
    Industrial wireless systems involving devices, actuators and robots that require ultra-reliable and low-latency communications (URLLC) is anticipated to grow in the future sixth generation of wireless communications (6G) \cite{She2021_tutorial,MTCwhitePaper2020}. Industrial Internet of things (IIoT) is the industrial application of IoT connectivity along with networking and cloud computing based on data analytic collected from IoT devices. Industrial environments are very diverse and heterogeneous as they are characterized by a large number of use-cases and applications. An underlying commonality among these diverse applications is that the wireless industrial automation connectivity solutions envisioned in Industry 4.0 (initialized in 5G) \cite{Aceto2019} will leverage cloud computing and machine learning throughout the manufacturing process. The expected URLLC key performance indicators (KPIs) in 6G networks are \textit{reliability} up to $1-10^{-9}$, \textit{latency} around $0.1\sim 1$ ms round-trip time, and \textit{jitter} in the order of $1$ $\mu s$ for industrial control networks \cite{MTCwhitePaper2020}. There is also high data rate demand due to increased number of sensors and their resolution, e.g., for robots.  In URLLC both the data and meta data sizes are small while both parts need to be very robust and have minimal error \cite{Mahmood2021ANetworks}. Thus, joint encoding of data and meta data is beneficial in terms of coding gain \cite{Popovski2019}. In addition, as the packet lengths in URLLC are usually small, the finite blocklength (FBL) theory is leveraged to investigate the achievable rate \cite{Polyanskiy2010}. 
    
    Reconfigurable intelligent surface (RIS) has been recognized as a promising technology to enhance the energy efficiency, and spectral efficiency of wireless communications \cite{DiRenzo2020}. An RIS is composed of meta-materials where the phase and amplitude of each element can be adjusted. This allows the reflected signal to have a desired effect, e.g., enhance the received signal-to-interference-plus-noise ratio (SINR) at a given location. Because of this feature, the distribution of the received signal, in the case of a blocked transmitter-receiver channel, has very little variation. The performance of such systems depend on the quantization levels at each phase shift element or circuitry impairments \cite{Hashemi2021a,Hashemi2021c}. Thus, the application of the RIS technology in industrial automation in ensuring high reliability is very promising \cite{Hashemi2021a}. Furthermore, since there is no processing overhead at the RIS  and the increase in the delay spread caused by an RIS is rather small, unlike conventional relays, URLLC latency requirements can be satisfied as well by a suitable design in higher layer.  Therefore, the RIS technology has high potential in URLLC applications. 
    
    There are a number of challenges when deploying RIS technology in practical industrial automation use cases. For instance, efficient physical layer design techniques, e.g., channel estimation, phase shift and amplitude response control and system-level optimizations, are still challenging and considered as active research topics. Toward this goal, optimization-oriented approaches relying on exhaustive alternating optimization methods have been introduced in the existing literature. Note that due to the unit modulus phase shifting constraint, the associated optimizations in the existing literature are highly non-convex and non-linear \cite{Wu2020c}. Thus, achieving a sub-optimal phase shift design is highly complicated and time-consuming. Additionally, since the radio channel characteristics vary over time or frequency, optimization-based methods need to be continuously tuned/re-executed to find the optimized phase shift values at the RIS which is impractical in mission-critical and sensitive industrial automation scenarios. Furthermore, the complexity of phase shift design optimizations increases considering the practical RIS in which the amplitude response changes by the value of phase shift in a non-linear manner \cite{Abeywickrama2020}. This poses new challenges to the existing optimization-based approaches which are still sophisticated and hard to solve even for ideal RISs \cite{Jinghe2021Aug}.

    In recent years, machine learning methods, particularly deep reinforcement learning (DRL) algorithms, have been considered as a reliable and powerful framework in wireless communications \cite{Jinghe2021Aug,Ali2020a}. The DRL methods rely on taking action and receiving a certain reward based on the action and interacting with the environment, which constructs the agent's experience. Thus, these methods usually do not require large training data set, which is highly beneficial in practical resource allocation problems in wireless communications. Therefore, the applicability of DRL toward more reliable and faster solutions in the next generations of URLLC is highlighted with the advent of efficient new algorithms \cite{Samad2019_magazine,Ali2020a,Bennis2018}. In this paper, our aim is to investigate practical phase shift design and optimization of a RIS-assisted URLLC system in an industrial automation by employing a novel and sophisticated DRL algorithm named as twin delayed deep deterministic policy gradient (TD3) \cite{fujimoto2018addressing}. 

	\subsection{Related Work}
    The resource allocation problems in RIS-assisted URLLC systems over short packet communications is a relatively new topic and have only been investigated in a few papers \cite{Ghanem2021WCNC,Ranjha2020,Xie2021Letter,AL-Mekhlafi2021}. In \cite{Ghanem2021WCNC} the authors studied an optimization problem for beamforming and phase shift control in a RIS-enabled orthogonal frequency division multiple access (OFDMA) URLLC system where the cooperation of a set of base stations (BSs)  to serve the URLLC traffic was discussed.  In \cite{Ranjha2020} the unmanned aerial vehicles (UAVs) trajectory and channel blocklength (CBL) in FBL regime as well as phase shift optimization in a RIS-aided network to minimize the total error probability was investigated. 
    In \cite{Xie2021Letter} a user grouping, CBL and the reflective beamforming optimization in a URLLC system was studied where a dedicated RIS assists the BS in transmitting  short packets in FBL scenario. The proposed optimization problem was tackled by semi-definite relaxation method and the user grouping problem is solved by a greedy algorithm. The authors in \cite{AL-Mekhlafi2021} studied the applicability of the RIS in joint multiplexing of enhanced mobile broadband (eMBB) and URLLC traffic to optimize the admitted URLLC packets while minimizing the eMBB rate loss to ensure the quality of service of the two traffic types by designing RIS phase shift matrices. It is worth noting that in all of the aforementioned works, the proposed problems were tackled by complex optimization-based algorithms as they usually are based on iterative algorithms. Particularly, even with an appropriate method that considers the non-linear amplitude response at the RIS elements, the computational complexity of such algorithms will still be significant. 
    
    Several existing works such as \cite{Yang2021,Feng2020b,Faisal2021,Huang2021,Huang2021a,Huang2020b,Zhang2021WCSP,zhu2022deep,Liu2022, Lin2020b,Feriani2021,Gong2022,Yang2021b, Liu2020d,Guo2021,Nguyen2021IntelligentLearning,Jiao2021} elaborated recent advances in DRL techniques on phase shift design at the RIS. In \cite{Yang2021} the secrecy rate of a wireless channel with RIS technology was maximized with quality of service (QoS) constraints on the secrecy rate and data rate requirements of the users. The resulting problem is solved by a novel DRL algorithm based on post-decision state and prioritized experience replay methods. 
    In \cite{Feng2020b} deep deterministic policy gradient (DDPG) method was employed to maximize the received signal-to-noise ratio (SNR) in a downlink multiple-input single-output (MISO) system via adjusting the phase shifts at the RIS. The authors in \cite{Faisal2021} discussed and compared the half-duplex and full-duplex operating modes in a RIS-aided MISO system and investigated the RIS reflective phase shift design via DDPG method. Joint relay selection and RIS reflection coefficient optimization in cooperative networks were studied in \cite{Huang2021}. The work in \cite{Huang2021a} considered maximizing the total achievable rate in infinite blocklength regime, i.e., assuming Shannon capacity, over a multi-hop multi-user RIS-aided wireless terahertz communication system. The authors in \cite{Huang2020b,Zhang2021WCSP,zhu2022deep} studied a RIS-assisted MISO system to adjust the BS transmit beamforming and the passive beamforming at the RIS via DDPG \cite{Huang2020b,Zhang2021WCSP} or soft actor-critic (SAC) \cite{zhu2022deep} methods. The objective of the considered problem is the total achievable rate in infinite blocklength regime across the network while considering ideal RIS \cite{Huang2020b,zhu2022deep} or practical RIS \cite{Zhang2021WCSP} with continuous phase shift model and the maximum transmit power of the BS. A recent study in \cite{Liu2022} investigated the applicability of distributed proximal policy optimization (PPO) technique in active/passive beamforming at the BS/RIS in a multi-user scenario. It is worth noting that the considered problem is defined in infinite CBL regime under Shannon rate formula and the optimization of CBL was not the topic of interest.

    On the other side, some studies such as \cite{Lin2020b,Feriani2021,Gong2022} discussed the utilization of model-free and model-based DRL algorithms in joint active/passive beamforming of RIS-assisted networks. In \cite{Lin2020b} the learning performance was improved by proposing a novel optimization-driven DDPG approach to optimize the RIS phase shift elements' values albeit at the cost of higher complexity. A comparison between DRL methods with optimization-based algorithms in RIS phase shift allocation was studied in \cite{Feriani2021,Gong2022}.  The application of DDPG algorithm in non-orthogonal multiple access (NOMA) networks employing RIS technology was also  discussed in \cite{Yang2021b}. Also, the studies in \cite{Liu2020d,Guo2021,Nguyen2021IntelligentLearning,Jiao2021} discussed the application of DRL methods in RIS-assisted UAV systems. The authors in \cite{Liu2020d} considered a downlink MISO system to adjust the RIS phase shifts as well as the coordinate of the flying UAV and transmit power via a decaying deep Q network (DQN) algorithm. The maximization of the millimeter wave (mmWave) secrecy rate by jointly optimizing the UAV trajectory, beamforming vectors and RIS phase shift was conducted in \cite{Guo2021} in which two independent DDPG networks, i.e., twin DDPG were leveraged to allocate the action strategies. The channels' feedback delay which results in channel coefficients' obsolescence, was also taken into account and the performance loss due to this effect was assessed. In \cite{Nguyen2021IntelligentLearning}, the DDPG algorithm was employed to optimize the power and the reflective surface phase shift design in a multi-UAV-enabled network while the authors in \cite{Jiao2021} analyzed the application of DDPG algorithm in UAV trajectory, power control and RIS phase shift optimization in a NOMA network.

	\subsection{Contributions}
    Despite the interesting results in the aforementioned works on joint active/passive beamforming design in RIS-aided communications, the optimization of the CBL and beamforming at the BS/RIS while considering the impact of impairments in practical RIS with non-linear amplitude response on the performance of a URLLC system over FBL regime has not been investigated before. In addition, most of the prior studies assumed that the RIS is ideal and the scenario is infinite blocklength regime while the conventional DDPG algorithm was utilized to solve the proposed resource allocation problem. However, several drawbacks are associated with this method, i.e., overestimation of the action-value function, unexpected actions and sudden performance degradation due to frequent policy network update which are addressed meticulously in the novel twin-delayed DDPG, i.e., TD3 method. 
    Motivated by the compelling works on resource allocation via DRL methods in RIS communications, we aim to elaborate the joint active/passive beamforming and CBL allocation problem where the objective is to maximize the total FBL rate subject to non-linear equality constraint for amplitude-phase response at the RIS. The contributions of our work are summarized in the following:
	\begin{itemize}
    	 \item A multi-antenna BS serving multiple actuators in the presence of a practical RIS is considered in industrial automation scenario. The RIS imperfections are modeled based on the empirical amplitude response in terms of phase shift values. The total FBL rate in downlink with arbitrary precoding at the BS, subject to a target block error probability (BLER), is considered as the system performance indicator over short packet communications. Based on the proposed system model, the CBL allocation and beamforming optimization at the BS and the RIS is formulated in which the objective is to maximize the total FBL rate of all actuators subject to non-linear amplitude responses at the RIS elements, total transmit power budget at the BS and total available CBL.
	    
	   \item Since the formulated problem is highly non-linear and non-convex, we invoke a novel policy gradient actor-critic DRL algorithm to solve the problem. Specifically, we leverage TD3 method that employs two individual deep neural networks (DNNs) to reduce the estimation error of action-values. Also, TD3 updates the main policy network less frequently than critic networks to stop overestimation of the action-value function, which usually leads to the policy breaking. 
	    
	   
	  \item The numerical results demonstrate that while the TD3 algorithm is well-suited to the proposed problem compared to typical SAC schemes, optimizing CBLs between actuators and performing active/passive beamforming design in the practical RIS systems with imperfections improves the network total FBL rate and reduces the transmission duration significantly. Furthermore, the performance reduction gap between an ideal RIS with continuous phase shift and the non-ideal RIS considering non-linear amplitude response is elaborated. Also, we show that by optimizing CBLs among actuators the transmission duration reduces by 17\% compared with equal CBL allocation.
    \end{itemize}

	\subsection{Notations and Structure of the Paper}
	In this paper, $\textbf{h} \sim \CN(\textbf{0}_{N\times 1},\textbf{C}_{N\times N})$ denotes a $N$-dimensional circularly-symmetric (central) complex normal distribution vector with N-dimensional zero mean vector $\textbf{0}$ and covariance matrix $\textbf{C}$. The operations $[\cdot]^\text{H}$, $[\cdot]^\text{T}$ denote the transpose and conjugate transpose of a matrix or vector, respectively. Also, the operators $\mathbb{E}[\cdot]$ and $\mathbb{V}[\cdot]$ denote the statistical expectation and variance, respectively.

	The structure of this paper is organized as follows. In Section \ref{SysModelSec}, the system model and the FBL rate is proposed, then the optimization framework of joint active/passive beamforming design and CBL allocation is presented. In Section \ref{DRLSec} the DRL preliminaries and exploited solution approach is studied. The numerical results are presented in Section \ref{NumericalSec}. Finally, Section \ref{ConslusionSec} concludes the paper.

	\section{System Model and Problem Formulation}
	\label{SysModelSec}
	\subsection{System Model}
\begin{figure}[t]
		\centering
		\includegraphics[scale=0.5]{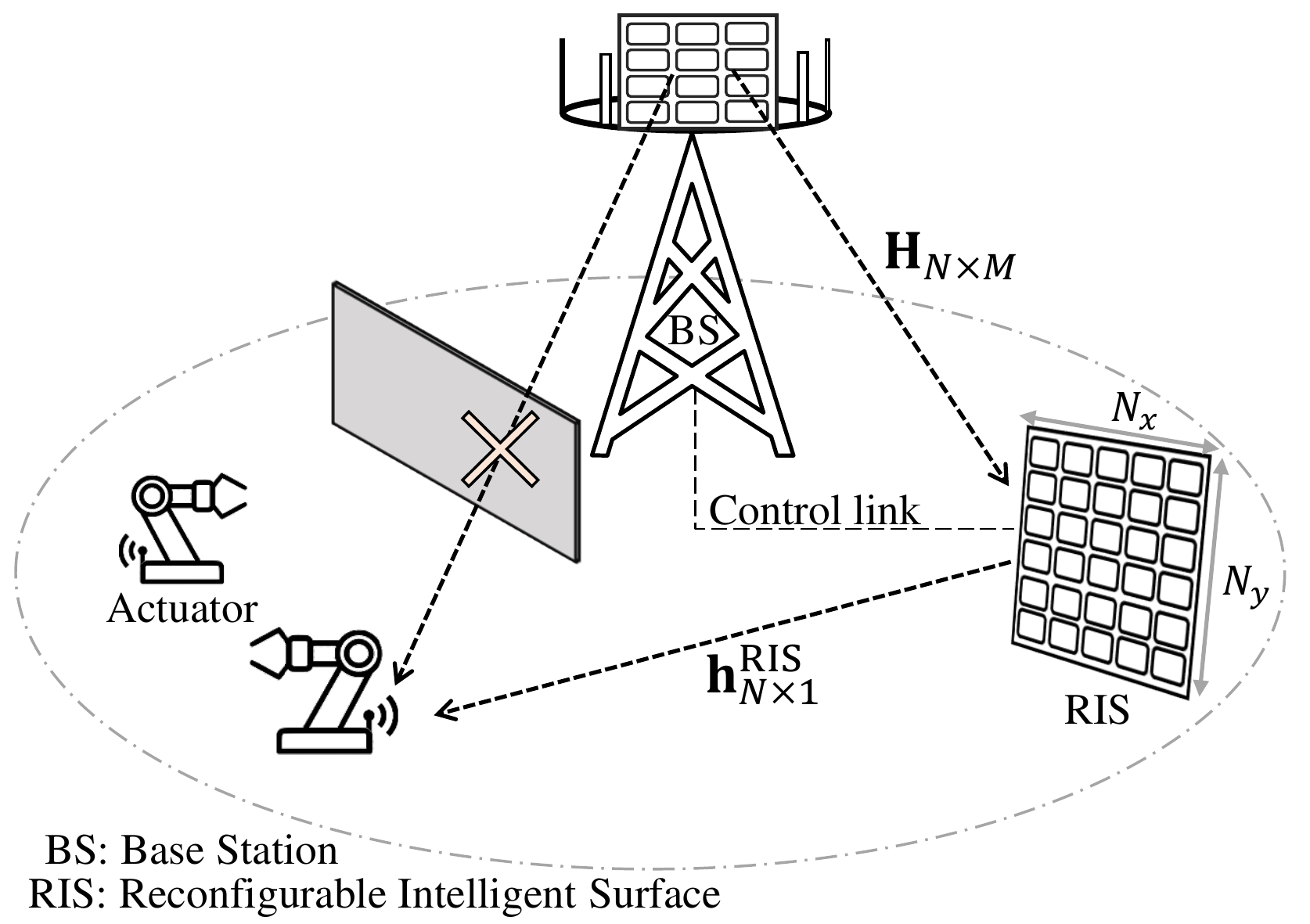}
		\caption{The considered system model.}
		\label{fig:0}
    \end{figure}

    Consider the downlink (DL) of an RIS-assisted wireless network in a factory setting which consists of a BS with $M=M_x\times M_y$ uniform planar array (UPA) antennas and $K$ single antenna actuators as illustrated in Fig. \ref{fig:0}. The RIS which has $N=N_x\times N_y$ phase shift elements constructs a communication channel between the actuators, and multi-antenna BS. It is assumed that the direct channels between the BS and actuators are blocked by possible obstacles in the factory and only reflected channels exist. Thus, the total channel response between the BS and an actuator is established by the reflected path from the RIS. The channel matrix $\textbf{H}\in \mathbb{C}^{N \times M}$ between BS and the RIS is denoted by  
    \begin{flalign}
    \textbf{H} = \sqrt{\frac{\zeta}{\zeta+1}}\overline{\textbf{H}}_{\text{LoS}}+\sqrt{\frac{1}{\zeta+1}}\textbf{H}_{\text{NLoS}}= [\textbf{h}^{\text{inc}}_1,...,\textbf{h}^{\text{inc}}_M],\label{H_model}
    \end{flalign}
    with the column vectors $\textbf{h}^{\text{inc}}_m=\sqrt{\frac{\zeta}{\zeta+1}}\overline{\textbf{h}}^{\text{inc}}_m +\sqrt{\frac{1}{\zeta+1}}\tilde{\textbf{h}}^{\text{inc}}_m$ for $\forall m \in \{1,...,M\}$ where each non-line-of-sight (NLoS) channel vector is distributed as $\tilde{\textbf{h}}^{\text{inc}}_m \sim \CN{(\textbf{0}_{M\times1},\beta^{\text{inc}}\textbf{I}_M)}$ in which  $\beta^{\text{inc}}$ is the pathloss from BS to the RIS, and $\textbf{I}_M$ is an identity matrix of size $M$. The proportion of line-of-sight (LoS) to the NLoS channel gain is defined as the Rician parameter $\zeta$. Additionally, the LoS channel $\overline{\textbf{H}}_{\text{LoS}}=[\overline{\textbf{h}}^{\text{inc}}_1,...,\overline{\textbf{h}}^{\text{inc}}_M]$ is defined as \cite{Jia2020a}
    \begin{flalign}
           \overline{\textbf{H}}_{\text{LoS}} = \sqrt{\beta^{\text{inc}}}\textbf{a}^{\text{H}}(\phi^{a}_1,\phi^{e}_1,N_x,N_y) \times \textbf{a}(\phi^{a}_2,\phi^{e}_2,M_x,M_y),
    \end{flalign}
    where $\phi_{1}^{a/e}$ denote the azimuth (elevation) angle of a row (column) of the UPA at the RIS and the projection of the transmit signal from BS to the RIS on the plane of the UPA at the RIS. Similarly, $\phi_{2}^{a/e}$ shows the azimuth (elevation) angle between the direction of a row (column) of the UPA at BS and the projection of the signal from BS to the RIS on the plane of the UPA at BS. In addition, the vector $\textbf{a}(x,y,N_1,N_2)$ is defined by \cite{Jia2020a}
    \begin{flalign}
    \textbf{a}(x,y,N_1,N_2)=\text{rvec}\left(\boldsymbol{\mathcal{H}}\right),
    \end{flalign}
    where rvec($\cdot$) denotes the row vectorization of a matrix, and 
    \begin{flalign}
      \boldsymbol{\mathcal{H}}=\left(e^{\mathrm{j} \mathcal{G}(x,y,n_1,n_2)}\right)_{n_1=1,2,...,N_1, n_2=1,2,...,N_2} \in \mathbb{C}^{N_1 \times N_2}  
    \end{flalign}
    such that each element row $n_1$ and column $n_2$ is constructed by means of \cite{Jia2020a}
    \begin{flalign}
    \mathcal{G}(x,y,n_1,n_2)=2\pi \frac{d}{\lambda}\left[(n_1-1)\cos x+(n_2-1)\sin x\right]\sin y,
    \end{flalign}
    in which $\lambda$ is the operating wavelength, and $d\leq \frac{\lambda}{2}$ is the antenna/element spacing.
    Similarly, the channel between RIS and actuator $k$ is 
    \begin{flalign}
        \textbf{h}_{k}^{\text{RIS}} = \sqrt{\frac{\zeta_k^{\text{RIS}}}{\zeta_k^{\text{RIS}}+1}}\overline{\textbf{h}}_{k}^{\text{RIS}} + \sqrt{\frac{1}{\zeta_k^{\text{RIS}}+1}}\tilde{\textbf{h}}_{k}^{\text{RIS}},
        \label{Rician_model}
    \end{flalign}
    where the Rician parameter $\zeta_k^{\text{RIS}}$ controls the proportion of LoS to the NLoS channel gain in actuator $k$. The NLoS channel is distributed as $\tilde{\textbf{h}}_{k}^{\text{RIS}} \sim \CN(\textbf{0}_{N\times1},\beta^{\text{RIS}}_k\textbf{I}_N)$ such that $\beta^{\text{RIS}}_k$ is the pathloss coefficient from RIS to actuator $k$. Furthermore, the LoS channel $\overline{\textbf{h}}_{k}^{\text{RIS}} \in \mathbb{C}^{N\times 1}$ is modeled by
    \begin{flalign}
        \overline{\textbf{h}}_{k}^{\text{RIS}}=\sqrt{\beta^{\text{RIS}}_k}\textbf{a}(\phi^{a,k}_3,\phi^{e,k}_3,N_x,N_y),   \quad \forall k \in \mathcal{K},
    \end{flalign}
    in which $\mathcal{K} = \{1,2,\dots,K\}$, and $\phi^{a,k}_3, \phi^{e,k}_3$ are the azimuth/elevation angles between RIS and the actuator $k$. 
    
	In this work we assume single-shot transmissions, i.e., retransmissions are not considered \cite{MPJ+16_oneStage,Anand2018}. Thus, the transmission latency is equal to one transmission time interval, which can be as low as $\sim0.1$ ms when adopting the flexible numerology introduced in 5G New Radio \cite{LatencyRef}. This assumption allows us to investigate the lower-bound performance of the proposed URLLC system as retransmissions improve the reliability of the system while at the cost of increasing latency \cite{Popovski.2018}. Nevertheless, some studies have compared the retransmission schemes with single-shot transmission \cite{Anand2018, Avranas2018}. As an example, the study in \cite{Avranas2018} employed incremental redundancy hybrid automatic repeat request (IR-HARQ) and concluded that the energy saving of the system enhances in comparison with single-shot transmission.

	For the considered system model, the received signal at the actuator $k$ in time instance $t$ is
	\begin{flalign}
	    \label{received_sig}
	    y_k[t] =  &\overset{\text{Actuator $k$ signal}}{\overbrace{\left( {\textbf{h}_k^{\text{RIS}}}^\text{H}\boldsymbol{\Theta}\textbf{H}\right)\boldsymbol{\omega}_k x_k[t]}} \\ \nonumber & + \underset{\text{Interference plus noise}}{\underbrace{ \left({\textbf{h}_k^{\text{RIS}}}^\text{H}\boldsymbol{\Theta}\textbf{H}\right) \sum_{i=1,i\neq k}^{K} \boldsymbol{\omega}_i x_i[t] +  n_k[t]}},
	\end{flalign}
	where $\boldsymbol{\omega}_k \in \mathbb{C}^{N\times 1}$ is the beamforming vector applied at the transmitter to the symbol $x_k[\cdot]$ of actuator $k$ with $\mathbb{E}[|x_k|^2]=1$. Also, $\lVert\boldsymbol{\omega}_k\rVert^2_2 = p_k$ in which $p_k$ is the transmit power allocated for actuator $k$ such that $\sum_{k=1}^{K}p_k=p_{\text{total}}$ is the BS transmit power, and $n_k[t]$ is the additive white Gaussian noise with $\mathbb{E}[|n_k[t]|^2] = N_0 W=\sigma^2$ where $N_0$, $W$ are the noise spectral density and the system bandwidth, respectively. The complex reconfiguration matrix $\boldsymbol{\Theta}_{N \times N}$ indicates the phase shift setting of the RIS which is defined as 
	\begin{flalign}
	    \boldsymbol{\Theta}_{N \times N} = & \text{ diag}(\beta_1 e^{j\theta_1}, \beta_2 e^{j\theta_2},..., \beta_N e^{j\theta_N}), \nonumber \\ 
	    & \beta_n \in  [0,1], \quad
	    \theta_n \in [-\pi,\pi), \quad \forall n \in \mathcal{N}
	\end{flalign}
	where $\mathcal{N} = \{1,2,...,N\}$. Note that in our model we have assumed that the RIS elements have no coupling or there is no joint processing among elements \cite{DiRenzo2020}. However, practical RIS phase shifters have phase-dependent amplitude response which is given by \cite{Abeywickrama2020}
	\begin{flalign}
           \beta_n(\theta_n) = (1-\beta_{\text{min}})\left(\frac{\sin(\theta_n-\phi)+1}{2}\right)^{\alpha}+\beta_{\text{min}},\label{practical_phase}
	\end{flalign}
	where $\beta_{\text{min}} \geq0$ (minimum amplitude), $\alpha \geq0$ (the steepness) and $\phi \geq0$ (the horizontal distance between $-\frac{\pi}{2}$ and $\beta_{\text{min}}$) are circuit implementation parameters. Note that, $\beta_{\text{min}} =1$ results in an ideal phase shifter.

    Based on the received signal at actuator $k$ in \eqref{received_sig}, the corresponding SINR achieved at time instance $t$ is given by
	\begin{flalign}
        \text{SINR}_k = \frac{\left| {\textbf{h}_k^{\text{RIS}}}^\text{H}\boldsymbol{\Theta}\textbf{H}\boldsymbol{\omega}_k\right|^2}{\displaystyle\sum_{i=1,i\neq k}^{K} \left| {\textbf{h}_k^{\text{RIS}}}^\text{H}\boldsymbol{\Theta}\textbf{H} \boldsymbol{\omega}_i\right|^2 + \sigma^2},\label{SINR_expression}
    \end{flalign}
    to cast the channel coefficients into one single matrix, and defining $\boldsymbol{\theta}=[\beta_1 e^{j\theta_1}, \beta_2 e^{j\theta_2},..., \beta_N e^{j\theta_N}]^{\text{H}}$ the SINR expression in \eqref{SINR_expression} can be rewritten as
    \begin{flalign}
      \text{SINR}_k = \frac{\left| \boldsymbol{\theta}^{\text{H}}{\tilde{\textbf{H}}_k}\boldsymbol{\omega}_k\right|^2}{\displaystyle\sum_{i=1,i\neq k}^{K} \left| \boldsymbol{\theta}^{\text{H}}{\tilde{\textbf{H}}_k} \boldsymbol{\omega}_i\right|^2 + \sigma^2},
    \end{flalign}
    where ${\tilde{\textbf{H}}_k}=\text{diag}({\textbf{h}_k^{\text{RIS}}}^\text{H})\textbf{H}$ and $\text{diag}(\cdot)$ refers to constructing a diagonal matrix based on a vector input as the diagonal elements. Herein, we concatenate the beamforming vectors such that $\bar{\boldsymbol{\omega}}=[\boldsymbol{\omega}_1,\boldsymbol{\omega}_2,...,\boldsymbol{\omega}_K]\in \mathbb{C}^{N\times K}$. According to the FBL theory, the number of information bits that can be transmitted through $c_k$ channel uses  over a quasi-static additive white Gaussian (AWGN) channel is given by \cite{Polyanskiy2010}
	\begin{flalign}
        L_k =  c_k\text{C}(\text{SINR}_k) - \mathcal{Q}^{-1}(\varepsilon_k)\sqrt{c_k\text{V}(\text{SINR}_k)} + \log_2(c_k),
         \label{achievable_rate_urllc}
    \end{flalign}
    where $\text{C}(\text{SINR}) = \log_2(1+\text{SINR})$ is the Shannon capacity which is defined in infinite blocklength regime and $\varepsilon_k$ is the target error probability for actuator $k$ while $\mathcal{Q}^{-1}(.)$ is the inverse of Q-function defined as $\mathcal{Q}(x) = \frac{1}{\sqrt{2\pi}}\int_{x}^{\infty}e^{-\nu^2/2}d\nu$. The channel dispersion is defined as
    \begin{flalign}
        \text{V}(\text{SINR}_k) = \frac{1}{(\ln2)^2} \left( 1- \frac{1}{(1+\text{SINR}_k)^2} \right),
    \end{flalign}
    Solving \eqref{achievable_rate_urllc} in order to find the decoding error probability $\varepsilon_k$ at the actuator $k$ yields
	\begin{flalign}
	    \varepsilon_k = \mathcal{Q} \left(f(\text{SINR}_k,c_k,L_k)\right),
	\end{flalign}
	where 
	\begin{flalign}
         f(\text{SINR}_k,c_k,L_k) = \sqrt{\frac{c_k}{V(\text{SINR}_k)}}(\log_2(1+\text{SINR}_k)-\frac{L}{c_k}).
	\end{flalign}
	Also, note that from \eqref{achievable_rate_urllc} when the blocklength $c_k$ asymptotically goes infinity the achievable rate simplifies to the conventional Shannon capacity formula. 

    \subsection{Problem Formulation}
    \label{ProbFormulationSec}
    Optimizing the total FBL rate of the actuators while ensuring the transmission target error probability by configuring the phase matrix of the RIS, beamforming matrix at the BS under optimized CBL vector $\textbf{c}=[c_1,c_2,...,c_K]$ is essential in factory environments to meet URLLC stringent requirements. Towards this goal, we formulate the following optimization problem (OP):
	\begin{flalign} \nonumber
		\textbf{P1 }& \max_{\bar{\boldsymbol{\omega}},\boldsymbol{\theta},\textbf{c}} \enskip   L_{\text{tot}} = \sum_{k=1}^{K}\left[\mathcal{V}_k(\bar{\boldsymbol{\omega}},\boldsymbol{\theta},\textbf{c}) - \mathcal{Q}^{-1}(\varepsilon_k^{\text{th}})\mathcal{W}_k(\bar{\boldsymbol{\omega}},\boldsymbol{\theta},\textbf{c})\right]
		\\ 
		\text{\textbf{s.t.}} \enskip 
		& \text{C$_\text{1}$: } \theta_n \in [-\pi,\pi), \enskip \forall n \in \mathscr{N}, \nonumber  \\
		& \text{C$_\text{2}$: }
		\beta_n = (1-\beta_{\text{min}})\left(\frac{\sin(\theta_n-\phi)+1}{2}\right)^{\alpha}+\beta_{\text{min}}, \enskip \forall n \in \mathscr{N},  \nonumber \\
		& \text{C$_\text{3}$: }
		\sum_{k=1}^{K} ||\boldsymbol{\omega}_k||^2_2 \leq p_{\text{total}}, \nonumber \\
		& \text{C$_\text{4}$: }
		\sum_{k=1}^{K} c_k \leq C, \enskip  c_k \geq c^{\text{min}}_k, \quad \forall k \in \mathcal{K}, \nonumber 
	\end{flalign} 
   where $\mathcal{V}_k(\bar{\boldsymbol{\omega}},\boldsymbol{\theta},\textbf{c}) = c_k\text{C}(\text{SINR}_k) + \log_2(c_k)$, and $\mathcal{W}_k(\bar{\boldsymbol{\omega}},\boldsymbol{\theta},\textbf{c})=\sqrt{c_k\text{V}(\text{SINR}_k)}$. The objective is to maximize the total number of information bits across all actuators and the variables are the reflective phase shift values of each element in $\boldsymbol{\theta}$ at the RIS. The aim of transmission in the FBL regime is to ensure the BLER at a target value which is equal to $\varepsilon_k^{\text{th}}$ $\forall k \in \mathcal{K}$ in the objective function. Thus, by maximizing the objective in \textbf{P1} while transmitting with the specified FBL rate, the target error probability can be ensured. The constraint C$_\text{1}$ denotes that the phase adjustment variable is chosen from the specified interval. C$_\text{2}$ implies the practical phase shift model which affects the amplitude response of the RIS. The maximum transmit power at BS is expressed in C$_\text{3}$. Also, C$_\text{4}$ is the constraint for total available number of CBLs at each transmission interval which is limited to maximum value of $C$. In addition, the CBL variable for each actuator $k$ must be at least $c_k^{\text{min}}$ so that the FBL regime rate is valid. It is observed from \textbf{P1} that it belongs to a class of nonlinear optimization problem which is thoroughly challenging to solve due to presence of equality constraint C$_\text{2}$. It is rational to use DRL for such problems since in DRL, the solution to the problem is the output of the forward pass to the neural network, which is a computationally simple process since it is often a set of simple operations. Further, the training of the neural networks that is done in different steps is performed in the background. Once the training is completed, the neural networks are updated. Therefore, the process to find the optimized variables in our problems is the inference of the neural networks that can be done in real-time \cite{Feng2020b}. Such a real-time solution cannot be obtained using optimization methods. Consequently, we employ a model-free DRL algorithm based on the TD3 algorithm described in the following section.

 	\section{DRL-based Formulation}
 	\label{DRLSec}
 	\subsection{Review on the Preliminaries}
	The goal of the agent in reinforcement learning (RL) is to \textit{learn} to find an optimal policy that maps states to actions based on its interaction with the environment so that the accumulated discounted reward function over a long time is maximized. A state contains all useful information from the sequence of the observations, actions and rewards. This kind of problems are tackled by representing them as Markov decision process (MDP) framework. An MDP is characterized by $(\mathcal{S},\mathcal{A},\mathcal{R},\mathcal{P}_{s\rightarrow s'})$ in which $\mathcal{S}$ is the set of environment states, $\mathcal{A}$ denotes the set of possible actions, which for this case  is defined in terms of the RIS phase shift values, $\mathcal{R}$ is the reward function, and $\mathcal{P}_{s\rightarrow s'}$ is the transition probabilities from state $s$ to $s'$, $\forall s,s' \in \mathcal{S}$. Mathematically, a Markov property means that the probability of next state (future state) is independent of the past given the present state. In RL algorithms, the environment can be fully or partially observable. In a fully observable environment, the agent directly observes the environment \cite{szepesvari2010algorithms}. The aim of the agent is to find an optimal policy to maximize the accumulated and discounted reward function over time-steps, i.e., to find $\pi^*$ in which the set of states $\mathcal{S}$ is mapped into the set of actions $\mathcal{A}$ as $\pi^*: \mathcal{S}\rightarrow \mathcal{A}$. The optimal policy $\pi^*$ maximizes the action-value function defined as
	\begin{flalign}
	   Q_{\pi}(s,a) = \mathbb{E}_{\pi}\left[\sum_{t=0}^{\infty}\gamma ^t r_{t+k+1}|S_t=s,A_t=a\right],
	   \label{Qfunction}
	\end{flalign}
	where the variable $0\leq \gamma  \leq 1$ is the discount factor to uncertainty of future rewards, $r_i$ is the acquired reward in step $i$ and $\mathbb{E}_{\pi}[\cdot]$ denotes the expectation with respect to policy $\pi$. By invoking Markov property and Bellman equation, \eqref{Qfunction} will be reformulated into,
	\begin{flalign}
	     Q_{\pi}(s,a) = \mathbb{E}_{\pi}\left[r_{t+1}+\gamma \sum_{a'\in\mathcal{A}}\pi(a'|s')Q_{\pi}(s',a')|S_t=s,A_t=a\right],
	\end{flalign}
    which $\pi(a'|s')$ gives the probability of choosing action $a'$ given that the agent is in state $s'$, the optimal value for action-value function can be achieved by \cite{sutton2018reinforcement}
    \begin{flalign}
           Q_{\pi^*}(s,a) =  \sum_{s' \in \mathcal{S},r\in \mathcal{R}}\text{Pr}(s'|s,a)\left(r+\gamma  \max_{a'}Q_{\pi^*}(s',a') \right),\label{OptimalQfunction}
    \end{flalign}
	where $\text{Pr}(s'|s,a)$ is the probability of transition to state $s'$ given that the agent is in state $s$ and the taken action is $a$. In order to find the optimal policy in \eqref{OptimalQfunction}, one must have knowledge about the transition probabilities that are usually unknown due to the environment structure.

	An RL problem can be solved by directly estimating the policy function rather than investigating the action-value function. These methods are named as policy-gradient (PG). In PG methods, the policy can be parameterized as a function of states on a DNN network. Let us denote the policy DNN with parameters' set  $\boldsymbol{\xi}^{\text{act}}$ as
	\begin{flalign}
       \pi(a|s;\boldsymbol{\xi}^{\text{act}}) = \text{Pr}(A=a|S=s;\boldsymbol{\xi}^{\text{act}}),
	\end{flalign}
	where $A=a$ is the action to be taken in state $S=s$. The probability of transiting to the state $s'$ from $s$ while taking action $a$ is shown as $\text{Pr}(s'|s,a)$. In PG methods the DNN weights are updated based on the result from policy-gradient theorem \cite{sutton2018reinforcement} which expresses that evaluating the gradient of the objective function given by
	\begin{flalign}\label{J_PG}
	   J(\boldsymbol{\xi}^{\text{act}}) \bigtriangleq \sum_{s \in \mathcal{S}}d^{\pi}(s)\sum_{a \in \mathcal{A}}\pi(a|s;\boldsymbol{\xi}^{\text{act}})Q(s,a;\boldsymbol{\xi}^{\text{crit}}),
	\end{flalign}
	is independent of the stationary distribution for states denoted as $d^{\pi}(s)$ for policy $\pi(\cdot)$. In \eqref{J_PG}, $Q(s,a;\boldsymbol{\xi}^{\text{crit}})$ represents the action-value function parameterized by $\boldsymbol{\xi}^{\text{crit}}$. The actor-critic networks are temporal difference (TD) learning methods that represent the policy function independent of the action-value function. We aim to employ actor-critic method where the policy is referred to as the actor that proposes a set of possible actions on a state. In actor-critic methods another DNN is employed to estimate the action-value function $Q(s,a;\boldsymbol{\xi}^{\text{crit}})$. The DNN evaluates the action-value by importing the current state and the action given by the policy network and its weights are represented as $\boldsymbol{\xi}^{\text{crit}}$.

	One of the efficient model-free and off-policy actor-critic methods that deals with the continuous action-space is DDPG \cite{lillicrap2015continuous}. In this algorithm, four DNNs are employed, two of them are for actor-critic networks and the other two are called target networks. The actor network directly gives the action by importing the states through a DNN with parameter set $\boldsymbol{\xi}^{\text{act}}$, i.e., $a=\mu(s;\boldsymbol{\xi^{\text{act}}})$ where $\mu(\cdot)$ denotes the deterministic policy meaning that the output is a value instead of a distribution. The critic network that has a DNN with $\boldsymbol{\xi}^{\text{crit}}$ weights evaluates the action-value function based on the action given by the policy network and the current state. The other two networks which are named as target networks give the target action-values in order to minimize the mean-squared Bellman error (MSBE) which is defined as \cite{szepesvari2010algorithms}
	\begin{flalign}\label{MSBE}
	   \mathcal{L}(\boldsymbol{\xi}^{\text{crit}},\mathcal{B}) \bigtriangleq \mathbb{E}\bigg[\bigg(Q(s,a;\boldsymbol{\xi}^{\text{crit}}) - (\overset{\text{target value}}{\overbrace{r+\gamma \max_{a'}Q(s',a';\boldsymbol{\xi}^{\text{crit}})}})\bigg)^2\bigg],
	\end{flalign}
	where the expectation is performed over ${(s,a,s',r)\sim \mathcal{B}}$ in which $\mathcal{B}$ is the experience replay memory which has stored the set of states, actions, rewards and the next states as a tuple $(s,a,r,s')$ over previous steps.  From \eqref{MSBE} the next optimal action $a'$ is calculated by the target actor network with parameter set $\boldsymbol{\xi}^{\text{targ-act}}$ where $a'=\mu(s';\boldsymbol{\xi}^{\text{targ-act}})$ and the corresponding action-value $Q(s',a';\boldsymbol{\xi}^{\text{targ-crit}})$ is then evaluated using the target critic network with weights $\boldsymbol{\xi}^{\text{targ-crit}}$. The two networks weights are usually copied over from the main network every some-fixed-number of steps by polyak averaging which is
	\begin{flalign}
	   \boldsymbol{\xi}^{\text{targ-act}} \leftarrow \tau\boldsymbol{\xi}^{\text{act}} +  (1-\tau)\boldsymbol{\xi}^{\text{targ-act}},\\
	   \boldsymbol{\xi}^{\text{targ-crit}} \leftarrow \tau\boldsymbol{\xi}^{\text{crit}} + (1-\tau)\boldsymbol{\xi}^{\text{targ-crit}},
	\end{flalign}
	where $\tau << 1$ is the hyperparameter used to control the updating procedure. 
	
	\subsection{Twin Delayed DDPG (TD3)}
\begin{figure*}[t]
     	\centering
     	\includegraphics[scale=0.80]{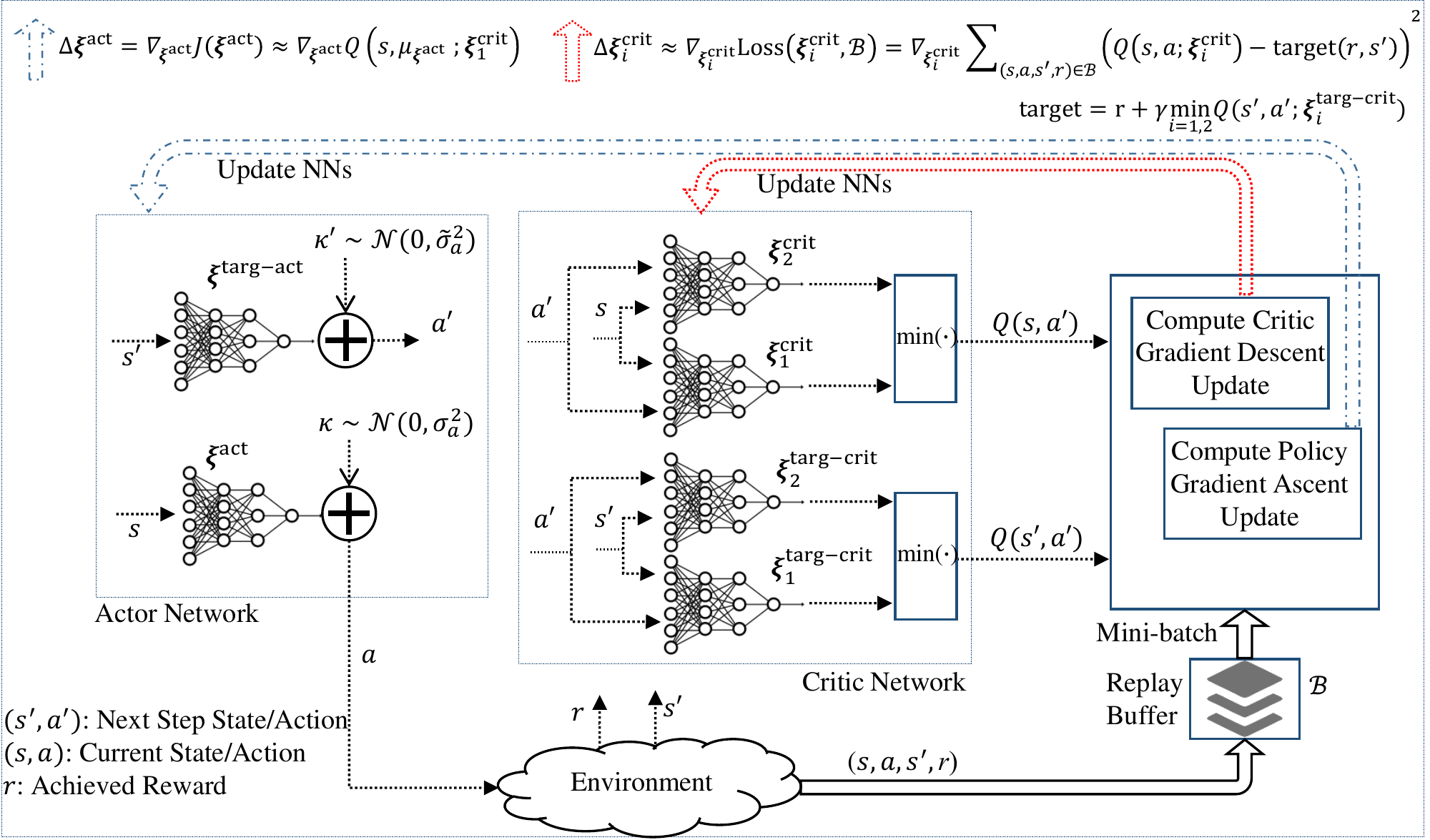}
     	\caption{The agent diagram of TD3 method.}
     	\label{fig:2_TD3}
    \end{figure*}

	Before proceeding with TD3 method, we restate the following Lemma from \cite{fujimoto2018addressing}:
	\begin{lemma}
	    \label{TD3BiasLemma}
	    For the true underlying action-value function which is not known during the learning process, i.e., $Q_{\pi}(s,a)$ and the estimated $Q(s,a;\boldsymbol{\xi}^{\text{crit}})$ the following inequality holds
	    \begin{flalign}
            \mathbb{E}\left[Q\left(s,a=\mu(s;\boldsymbol{\xi^{\text{act}}});\boldsymbol{\xi}^{\text{crit}}\right)\right] \geq \mathbb{E}\left[Q_{\pi}\left(s,a=\mu(s;\boldsymbol{\xi^{\text{act}}})\right)\right],
	    \end{flalign}
	\end{lemma}
	Based on Lemma \ref{TD3BiasLemma}, since the DDPG algorithm leverages the typical Q-learning methods, it overestimates the Q-values during the training which propagates throughout the next states and episodes. This effect deteriorates the policy network as it utilizes the Q-values to update its weights and hyperparameters and results in poor policy updates. The impact of this overestimation bias is even problematic with feedback loop that exists in DRL methods where suboptimal actions might be highly rated by biased suboptimal critic networks. Thus, the suboptimal actions will be reinforced in next policy updates.  The TD3 algorithm introduces the following assumptions to address the challenges \cite{fujimoto2018addressing}
	\begin{itemize}
	    \item As illustrated in Fig. \ref{fig:2_TD3}, TD3 recruits two DNNs for estimating the action-value function in the Bellman equation, then the minimum value of the output of Q-values is used in \eqref{MSBE}.
	    \item In this method, the target and policy networks are being updated less frequently than critic networks.
	    \item A regularization of the actions that can incur high peaks and failure to the Q-value in DDPG method is leveraged so that the policy network will not try these actions in the next states. Therefore, the action will be chosen based on adding a small amount of clipped random noise to the selected action as given by
         \begin{flalign}
            a'=\text{clip}(\mu(s';\boldsymbol{\xi}^{\text{targ-act}})+\text{clip}(\kappa',-c,+c),a_{\text{Low}},a_{\text{High}}),
		\end{flalign}
		where $\kappa' \sim \mathcal{N}(0,\tilde{\sigma}^2_a)$ is the added normal Gaussian noise and $a_{\text{Low}}$, $a_{\text{High}}$ are the lower and upper limit value for the selected action that is clipped to ensure a feasible action which may not be in the determined interval  due to added noise. Also, the constant $c$ truncates the added noise at inner stage to keep the target action close to the original action. 
        \end{itemize}

    The detailed description of the TD3 is given in Algorithm \ref{TD3Algo}. A central controller at the BS is collecting and processing the required information for the algorithm execution. First, the six DNNs are initialized by their parameter weights, i.e., the actor network $\boldsymbol{\xi}^{\text{act}}$, the critic networks $\boldsymbol{\xi}^{\text{crit}}_{i}$, $i \in \{1,2\}$ coefficients are initialized randomly while the target actor and critic networks' parameters are determined by replicating the primary actor and critic networks' coefficients, respectively. Also, the empty experience replay memory with specified capacity is prepared and the discount factor $\gamma$, learning rates, soft update hyperparameter $\tau$, maximum step size $N_{\text{steps}}$ and episodes $N_{\text{episode}}$ are determined. In the training stage, the reflective phase matrix at the RIS is randomly initialized. The current channel coefficients of the actuators is acquired and the state set is formed, correspondingly. Next, the action, i.e., the phase shift matrix is collected from the output of the actor DNN with parameter set $\boldsymbol{\xi}^{\text{act}}$ by importing the current state vector as the input. Next, the observed reward, taken action, the current state $s$, and the next state $s'$, i.e., the modified channels' coefficients in terms of the phase shift values given by the actor network are recorded at the experience replay buffer. To update the DNNs, a mini-batch of stored experience memory is randomly selected, then, the target actions are computed via target actor DNN with weights $\boldsymbol{\xi}^{\text{targ-act}}$ and the target values are evaluated by selecting the minimum value of target critic DNNs' output which correspond to minimizing the loss function by performing gradient descent method. In addition, when it is time to update the actor and target networks, e.g., out of $t'$ steps where typically $t'=2$ (once in every two steps), the gradient ascent is employed to compute the new coefficients of DNNs, i.e., renewal of $\boldsymbol{\xi}^{\text{targ-act}}$, $\boldsymbol{\xi}^{\text{targ-crit}}$, and $\boldsymbol{\xi}^{\text{act}}$.

	\begin{algorithm}[t]
    	\small
		\SetAlgoLined 
		\KwIn{The number of actuators, the RIS amplitude-phase response model, position of the BS and actuators in 2D-plane.}
		
		\KwOut{Trained agent with DNNs' weight coefficients.}
		\textit{Initialization}: Initial values for weights $\boldsymbol{\xi}^{\text{act}}$, $\boldsymbol{\xi}^{\text{crit}}_1$ and $\boldsymbol{\xi}^{\text{crit}}_2$, empty replay memory $\mathcal{B}$. Let $\boldsymbol{\xi}^{\text{targ-act}} \leftarrow \boldsymbol{\xi}^{\text{act}}$,  $\boldsymbol{\xi}_1^{\text{targ-crit}} \leftarrow \boldsymbol{\xi}_1^{\text{crit}}$ and $\boldsymbol{\xi}_2^{\text{targ-crit}} \leftarrow \boldsymbol{\xi}_2^{\text{crit}}$, soft update coefficient $\tau$, the discount factor $\gamma$, the learning rates, the maximum steps $N_{\text{steps}}$, and maximum episodes $N_{\text{episode}}$\;
		\For{$\texttt{e}=1,2,...,N_{\text{episode}}$}{
		    Randomly initiate CBLs, and beamforming at RIS/BS\;
		    Collect current channel coefficients $\left\{\textbf{H},\textbf{h}_k^{\text{RIS}}, \forall k\right\}$\;
		    \For{$\texttt{t}=1,2,...,N_{\text{steps}}$}{
			Select action $a=\text{clip}(\mu(s;\boldsymbol{\xi}^{\text{act}}) + \kappa,a_{\text{Low}},a_{\text{High}})$, where $\kappa\sim \mathcal{N}(0,\sigma^2_a)$\;
			Perform the action $a$ selected above\;
			Observe next state $s'$ and the reward value $r$\;
			Store the tuple $(s,a,s',r)$ in the replay memory $\mathcal{B}$\;
			Sample a batch of tuple $\mathbb{B} \subset \mathcal{B}$ from experience replay memory\;
			Compute target actions given as $a'=\text{clip}(\mu(s';\boldsymbol{\xi}^{\text{targ-act}})+\text{clip}(\kappa',-c,+c),a_{\text{Low}},a_{\text{High}})$ where $\kappa'\sim \mathcal{N}(0,\tilde{\sigma}^2_a)$\;
			Compute the target value $ \texttt{target}(r,s') =r +\gamma  \min_{i\in \{1,2\}}Q(s',a';\boldsymbol{\xi}_i^{\text{targ-crit}})$\;
			Update the critic networks by performing gradient descent for $i\in \{1,2\}$ using
			\begin{flalign}
			    \frac{1}{|\mathbb{B}|}\nabla_{\boldsymbol{\xi}_i^{\text{crit}}}\displaystyle\sum_{(s,a,s',r)\in \mathbb{B}}\left(Q(s,a;\boldsymbol{\xi}_i^{\text{crit}}) - \texttt{target}(r,s')\right)^2, \nonumber 
			\end{flalign}\\
			\If{time to update policy network ($t$ mod $t'$)}{
			Update the policy network by performing gradient ascent with
			\begin{flalign}
			    \frac{1}{|\mathbb{B}|}\sum_{s\in \mathbb{B}}\nabla_{a}Q(s,a=\mu(s;\boldsymbol{\xi}^{\text{act}});\boldsymbol{\xi}_1^{\text{crit}})\nabla_{\boldsymbol{\xi}^{\text{act}}}\mu(s;\boldsymbol{\xi}^{\text{act}}),\nonumber 
			\end{flalign}\\
			Update the target networks with
			\begin{flalign}
        	   \boldsymbol{\xi}^{\text{targ-act}} & \leftarrow \tau\boldsymbol{\xi}^{\text{act}} +  (1-\tau)\boldsymbol{\xi}^{\text{targ-act}},\nonumber \\
        	   \boldsymbol{\xi}_i^{\text{targ-crit}} & \leftarrow \tau\boldsymbol{\xi}_i^{\text{crit}} + (1-\tau)\boldsymbol{\xi}_i^{\text{targ-crit}}, \enskip \text{for $i\in \{1,2\}.$} \nonumber 
        	\end{flalign}
			}

			
			}{}{}
			}
		\caption{Twin Delayed DDPG (TD3) Algorithm}\label{TD3Algo}
	\end{algorithm}

	\subsection{Applying TD3 to Solve \textbf{P1}}
	A preliminary step to solve the problem \textbf{P1} with TD3 is to map the components and properly define the algorithm states, actions and the reward function. In this section, we investigate them in detail as follows:
	\subsubsection{States} 
	The agent interacts with the environment to optimize the FBL rate performance while ensuring a target BLER. Hence, the agent only has knowledge about the local information about actuators, e.g., the channel coefficients. Consequently, the DRL agent state space is defined as the aggregation of the angle and magnitude components of the composite channel coefficients, previous step beamforming vectors, and interference terms. First, it is useful to denote the interference and the inner terms as 
	\begin{flalign}
	\boldsymbol{\Upsilon}^{k}_t & = \boldsymbol{\theta}^{\text{H}}(t-1){\tilde{\textbf{H}}_k}, \\
     \Upsilon^{kk'}_t & =  \boldsymbol{\theta}^{\text{H}}(t-1){\tilde{\textbf{H}}_k}\boldsymbol{\omega}_{k'}(t-1),
	\end{flalign}
    where $\boldsymbol{\Upsilon}^{k}_t \in \mathbb{C}^{N\times M}$ and $\Upsilon^{kk'}_t \in \mathbb{C}$. The current state $s_t$ is constructed as follows:
	\begin{flalign}
          & s_t    =  \{s_t^1,s_t^2,s_t^3, s_t^4\}, \label{state} \\ \nonumber 
          & s_t^1 =    \Big\{   \left| \Upsilon^{kk'}_t \right|, \angle{\Upsilon^{kk'}_t} \mid \forall k,k' \in \mathcal{K}\Big\}, \\ \nonumber 
          & s_t^2  = \left\{\lVert\boldsymbol{\Upsilon}^{k}_t\rVert_2, \lVert \boldsymbol{\omega}_k(t-1)\rVert_2, \angle{\boldsymbol{\Upsilon}^{k}_t},  \angle{\tilde{\textbf{H}}_k}, \angle{\boldsymbol{\omega}_k(t-1)} \mid \forall k \in \mathcal{K}\right\}, \\ \nonumber 
        &  s_t^3 =\left\{\theta_n(t-1) \mid \forall n \in \mathcal{N}\right\}, \\ \nonumber 
        & s_t^4 = r_{t-1},
	\end{flalign}
	where $\mathcal{K}=\{1,2,...,K\}$ and $\mathcal{N}=\{1,2,...,N\}$. Note that the operators $\angle{\textbf{X}}$ and $|\textbf{X}|$ denote the angle and magnitude of each complex element in $\textbf{X}$, respectively. The size of state space in \eqref{state} is determined based on $2 K^2$ interference terms in $s_t^1$, $KM(N+2)+2K$ active beamforming coefficients and composite channel response from BS to the actuators in $s_t^2$, and $N$ RIS reflection variables in $s_t^3$. Also, the previous reward achieved in the last step is considered as $s_t^4$ which will be defined in subsequent sections. Thus, the total size of the state space is given by $|s_t| = 2K(K+1)+(N+1)(KM+1) + KM$.
	
	\subsubsection{Actions}
	 The action is determined as the value of phase shift at each element ($\theta_n(t)$, $\forall n$) and the action set in time $t$ is given by
	 \begin{flalign}\label{actions_}
        a_t = \left\{c_k(t), \left|\boldsymbol{\omega}_k(t)\right|, \angle \boldsymbol{\omega}_k(t) \mid \forall k \in \mathcal{K} \right\} \bigcup \left\{\theta_n(t) \mid \forall n \in \mathcal{N}\right\},
	 \end{flalign}
	 such that each phase shift element value is chosen from the interval $\theta_n(t) \in [-\pi,\pi)$, $\forall n$ by multiplying the corresponding outputs of $tanh(\cdot)$ layer by $\pi$. Also, each beamforming vector is generated by producing complex numbers with separate magnitude values and angle components, then scaling the resultant vectors such that the total transmit power at the BS is satisfied, i.e., $\sum_{k=1}^{K}\lVert\boldsymbol{\omega}_k\rVert^2_2=p_{\text{total}}$. To construct the actions corresponding to the CBLs, $K$ elements of $tanh(\cdot)$ output layer in actor network are selected as
	 \begin{flalign}
         a^{\textbf{c}}_t=\{a_1^\textbf{c},a_2^\textbf{c},...,a_K^\textbf{c}\}, \label{action_BL}
	 \end{flalign}
	 where $-1\leq a_k^\textbf{c} \leq 1$ $\forall k$. Considering $\textbf{c}^{\text{min}}=[c^{\text{min}}_1,c^{\text{min}}_2,...,c^{\text{min}}_K]$ as the minimum CBL vector, the actions in \eqref{action_BL} are scaled as follows to construct $\textbf{c}(t)$:
	 \begin{flalign}\label{BL_}
	    \tilde{a}^{\textbf{c}}_t &  \leftarrow \frac{a^{\textbf{c}}_t+1.0}{2}, \\ \nonumber 
        \textbf{c}(t) & \leftarrow \frac{C-\textbf{c}^{\text{min}}}{\sum_{k=1}^{K}\tilde{a}^{\textbf{c}}_k + \zeta} \tilde{a}^{\textbf{c}}_t + \textbf{c}^{\text{min}}, 
	 \end{flalign}
	 where $\zeta << 1$ is a small value to avoid possible division by zero as $0\leq \tilde{a}^{\textbf{c}}_k\leq 1.0$, $\forall k$. Consequently, from \eqref{BL_} and the procedure to generate beamforming vectors, we can easily confirm that C$_\text{1}$--C$_\text{4}$ are satisfied. Finally, given \eqref{actions_}, the output size of the actor network will be $K + 2KM + N$. 
	 
	\subsubsection{Reward Function} 
	The objective function in \textbf{P1} has to be maximized over time-steps $t$, i.e., $L_{\text{tot}}$. In addition, as explained in the previous section, by scaling procedure of the raw actions, the constraints in \textbf{P1} can be met to produce feasible actions without reflecting their violation penalty into the reward function. Thus, the agent's reward function at each time-step $t$ is designed to be
	\begin{flalign}
	        r_t =  \sum_{k=1}^{K}&\Big[\mathcal{V}_k\left(\bar{\boldsymbol{\omega}}(t),\boldsymbol{\theta}(t),\textbf{c}(t)\right) \nonumber \\  & - \mathcal{Q}^{-1}(\varepsilon_k^{\text{th}})\mathcal{W}_k\left(\bar{\boldsymbol{\omega}}(t),\boldsymbol{\theta}(t),\textbf{c}(t)\right)\Big],
	\end{flalign}
	

    In the following, we discuss the convergence proof for TD3 algorithm in a finite MDP setting with discrete action-space referred as clipped double Q-learning. It is worth noting that generalization to continuous action and actor-critic networks is straightforward. First, given $Q^1$ and $Q^2$ as the action-value estimator functions, the best action is determined based on $a^*=\argmax_{a}Q^1(s',a)$. Also, the target value is found by Bellman equation as $y=r+\gamma\displaystyle\min\{Q^1(s',a^*),Q^2(s',a^*)\}$. In double Q-learning the action-value tables are updated as $Q^i(s,a)=Q^i(s,a)+\alpha_t(y-Q^i(s,a))$, $i \in \{1,2\}$. Given this knowledge, the following theorem investigates the conditions for the convergence of clipped double Q-learning \cite{fujimoto2018addressing}. 
    \begin{theorem}
        The clipped double Q-learning will theoretically converge to the optimal action-value function $Q^*$ with probability 1 if the following assumptions hold:
        \begin{enumerate}
            \item The MDP is of finite size and the action space is sampled infinite number of times.
            \item The discount factor should be $\gamma\in[0,1)$ and the Q-values are stored in a look-up table.
            \item The learning rate should meet $\alpha_t\in[0,1]$, $\sum_t\alpha_t=\infty$, $\sum_t\alpha_t^2\leq \infty$. 
            \item $Q^1$ and $Q^2$ receive an infinite number of updates and $\mathbb{V}[r(s,a)]\leq \infty$ $\forall s,a$. 
        \end{enumerate}
        \label{theo_conv}
    \end{theorem}
    Consequently, from the conditions in Theorem \ref{theo_conv}, we can ensure that to solve \textbf{P1} by utilizing TD3 method, with proper selection of the learning rates, discount factor, and finite variance of the reward function the algorithm will converge to the optimized policy $\pi^*$. Since, the reward function is the objective in \textbf{P1}, it is needed to verify that $\mathbb{V}[r_t(s_t,a_t)]\leq \infty$, therefore we have
    \begin{flalign}
            \mathbb{V}&\bigg[\overset{A}{\overbrace{\sum_{k=1}^{K}\mathcal{V}_k\left(\bar{\boldsymbol{\omega}}(t),\boldsymbol{\theta}(t),\textbf{c}(t)\right)}} \\ \nonumber & -\bigg( \overset{B}{\overbrace{\sum_{k=1}^{K}\mathcal{Q}^{-1}(\varepsilon_k^{\text{th}})\mathcal{W}_k(\bar{\boldsymbol{\omega}}(t),\boldsymbol{\theta}(t),\textbf{c}(t))}}\bigg)\bigg] \nonumber \\&  =\mathbb{V}[A-B]=\mathbb{V}[A]+\mathbb{V}[B]-\text{COV}[A,B],
    \end{flalign}
	given that the number of RIS elements is finite $N\leq \infty$, the BS has finite transmit power, and the CBL variables $c_k$ $\forall k$ are bounded, then, the SINR values will have finite variance $\mathbb{V}[\text{SINR}]\leq \infty$ \cite{Hashemi2021a} which results in finite reward variance, and concludes that $\mathbb{V}[A-B]\leq \infty$.
	
	\subsection{Complexity Analysis}
	In this section, we discuss the computational complexity of proposed TD3 to solve the \textbf{P1}. Let $n_L$ be the number of layers in each DNN and $z_l$ be the number of neurons in layer $l$. Then, in the training mode, the evaluation and update in one time-step is $\mathcal{O}\left(|\mathbb{B}| \times \sum_{l=1}^{n_L-1}z_lz_{l+1}  \right)$ \cite{Yang2021} where $|\mathbb{B}|$ denotes the size of batch tuple. Since the TD3 algorithm has a finite number of DNNs and it takes $N_{\text{episode}}\times N_{\text{steps}}$ iterations to complete the training phase in which $N_{\text{steps}}$ is the number of steps in each episode and $N_{\text{episode}}$ is the total number of episodes. Therefore, the total computational complexity will be $\mathcal{O}\left(|\mathbb{B}| N_{\text{episode}}N_{\text{steps}}  \sum_{l=1}^{n_L-1}z_l z_{l+1}  \right)$. 
	
	\section{Numerical Results}
	\label{NumericalSec}
		In this section, we numerically evaluate the considered joint active/passive beamforming and CBL allocation optimization problem by using the proposed DRL method. A generic channel model is chosen to obtain insights about the proposed approach’s performance trends independent of the operating frequency and employed channel model. Evaluations under specific channel models, including indoor factory scenarios and millimeter wave channels, is left for future studies. Since, the components and robots in an industrial automation are in almost fixed position, we considered four actuators in a factory environment located in 2D-plane coordinates at $[16,40]$, $[32,40]$, $[48,40]$ and $[64, 40]$ where a BS is positioned at $[0, 0]$ and the RIS is located at $[40, 0]$. The large scale path loss fading is modeled as $\text{PL(dB)}=\text{PL}_0-10\nu\log_{10}(D\text{[m]})$ where $\text{PL}_0=-30$ dB, $\nu=2.2$ is the path loss coefficient and $D$ is the distance between the transmitter and the receiver \cite{Feng2020b}. Table \ref{tab1} shows the summary of the selected parameters for the network components during simulations.
	\begin{table}[t]
		\caption{Simulation parameters.}
		\centering
		\begin{tabular}{ l  m{3cm} }
			\hline
			Parameter & Default value \\ \hline
			Number of actuators ($K$)  & 4  \\ 
			Number of BS antennas ($M$)  & 4  \\ 
			Number of RIS elements ($N$)  & 16  \\ 
			BS transmit power ($p_{\text{total}}$)  & 1.0 mW  \\
			Target error probability ($\varepsilon^{\text{th}}_k, \forall k$)  & $10^{-8}$  \\
			Receiver noise figure (NF)  & 3  dB \\
			Noise power density ($N_0$) & -174 dBm/Hz \\ 
			Total available CBL ($C$) & 100 \\ 
			Minimum CBL ($c_k^{\text{min}}$, $\forall k$) & 10 \\ 
			Bandwidth ($W$) &  0.1 MHz \\ 
			Rician factors ($\zeta$ and $\zeta_k^{\text{RIS}}$ $\forall k$)  & 10  \\ 
			BS height &  12.5 m \\
			BS location in 2D plane & [0, 0] m \\
			RIS position in 2D plane & [40, 0] m \\
			RIS phase shifter parameters & \shortstack[l]{ $\beta_{\text{min}}=0.4$ \\ $\alpha=1.9$ \\ $\phi=0.43\pi$} \\
			\hline
		\end{tabular}
		\label{tab1}
	\end{table}

	The learning rate in actor and critic networks of TD3 agent is set to $\alpha_t=10^{-4}$. The actor network DNN has three hidden dense layers with $[800,400,200]$ neurons. The activation functions in all hidden layers are considered as rectified linear unit $ReLU(\cdot)$ except for the last layer in which for the actor network is assumed to be $tanh(\cdot)$ to provide better gradient. Since the output of $tanh(\cdot)$ is limited to the interval $[-1,1]$, it might get saturated for large inputs in most of the times. To avoid such saturation of the actions in the output of the actor network, the input state and action in the architecture of the critic networks are first processed by two dense layers with 800 neurons, separately. The implication behind this is that the actor network is being updated in the direction suggested by the critic, thus proper estimation of Q-values is of paramount importance to avoid such occurrence. Next, the resultant outputs are added and are given to dense layers with size $[600,400]$ to estimate the current Q-value at final stage. Also, extensive simulations revealed that employing Layer Normalization \cite{ba2016layer} helps to prevent the action value saturation, thus, we used this normalization technique before activation functions in dense layers.

	The experience replay buffer capacity is $10000$ with batch size $64$ such that the samples are uniformly selected from the buffer data. Furthermore, the exploration noises $\kappa,\kappa'$ in TD3 actor networks are zero-mean normal random variables with variance $\sigma^2_a=0.1, \tilde{\sigma}^2_a=0.1$. The target actor/critic networks' soft update coefficient is $\tau = 0.005$. During the updating procedure, the policy network is being updated every $t'=4$ steps. In all of the episodic illustrations, the agent is being evaluated over 100 independent realizations of the network channels to assess its performance, i.e., the illustrations are generalized results over 100 realizations.

\begin{figure*}[t]
    \begin{subfigure}{0.5\linewidth}
     	\centering
   		\includegraphics[trim = 8.5cm 8.5cm 8.5cm 8.5cm,scale=0.6]{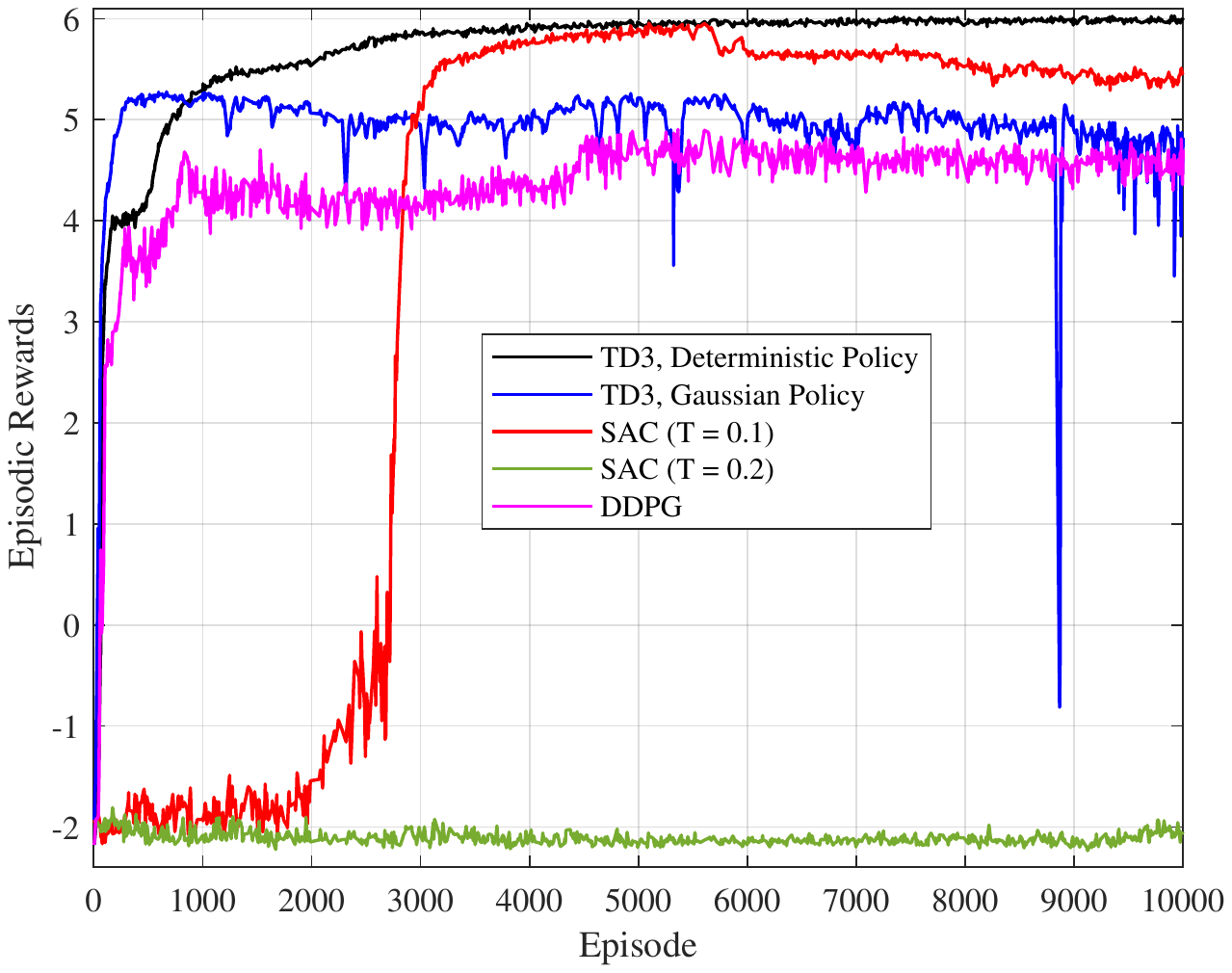}
     	\caption{ }
     	\label{fig:reward_vs_methods}
     \end{subfigure}
      \begin{subfigure}{0.5\linewidth}
     	\centering
   		\includegraphics[trim = 8.5cm 8.5cm 8.5cm 8.5cm,scale=0.6]{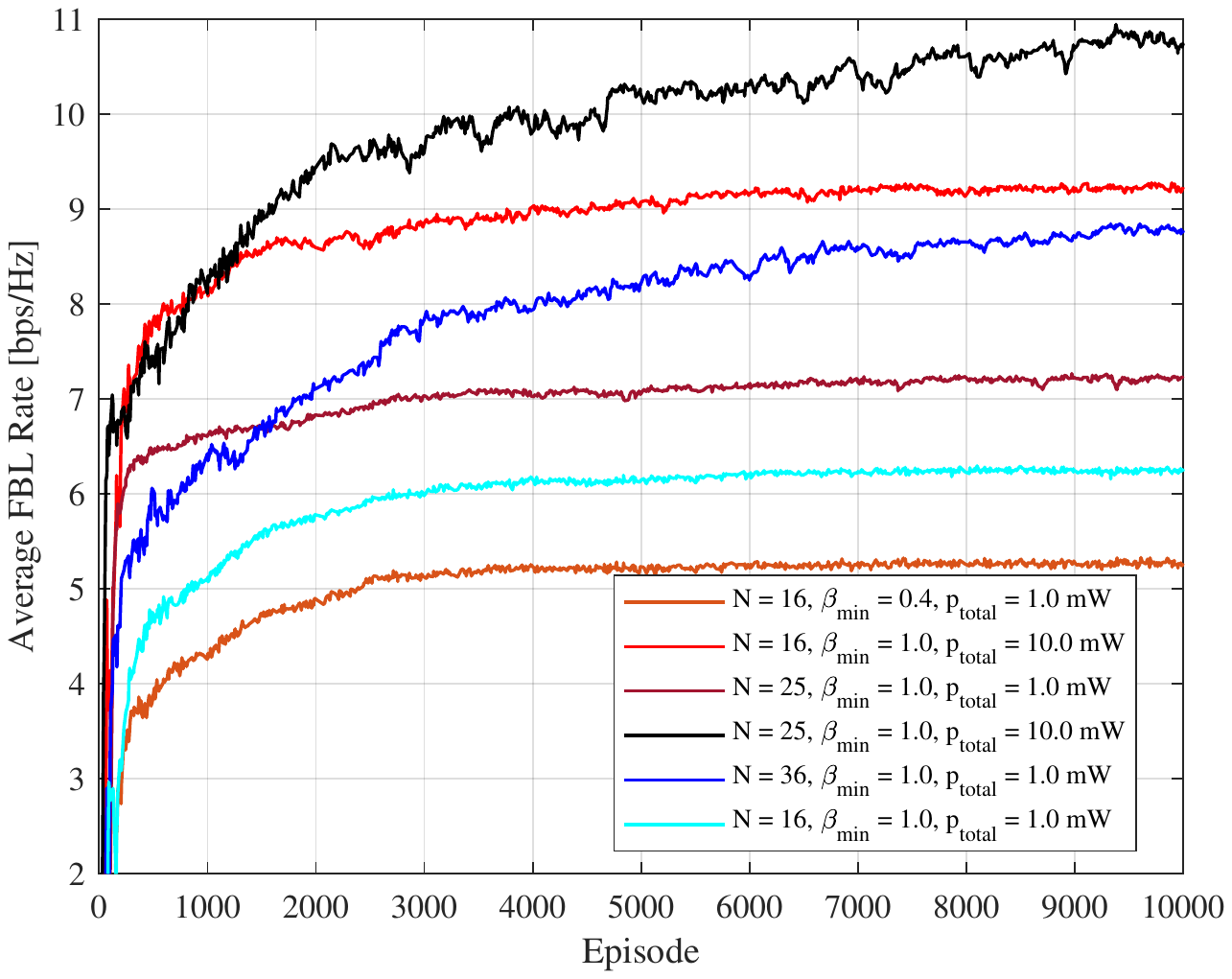}
     	\caption{ }
     	\label{fig:rewards}
     \end{subfigure}
      \caption{(a): The comparison between SAC and TD3 with Gaussian/Deterministic policies. (b): FBL rate behaviour versus episode for different number of elements at the RIS and BS transmit power budget.}
      \label{fig:convergence_fig}
    \end{figure*}

In Fig. \ref{fig:reward_vs_methods} the proposed TD3 method is compared with SAC algorithm with different entropy regularization coefficients ($T=0.1,0.2$) and DDPG. As observed, the DDPG has higher fluctuations in the curve of episodic average reward value compared to the TD3 algorithm. The fluctuations in DDPG method occurred due to frequent policy network updates and the overestimation bias which are eliminated in TD3. In addition, TD3 outperforms DDPG method in both final performance and learning speed in phase control.  It can be observed that the SAC with higher regularization value cannot learn the optimal policy corresponding to too much exploration, however, for lower values of the coefficient the agent started learning in around 3000 episodes, then the reward drops in around 5000 episodes. Also, the performance of employing Gaussian policy randomization at the output of actor network is illustrated as well as utilizing deterministic policy. Basically, in deterministic sampling the agent uses the mean action instead of a sample from fitting a Gaussian distribution with mean and variance dense layers. From illustrated curves it is perceived that deterministic policy outperforms randomized policy as the agent has reached higher reward value in deterministic policy method. In addition, employing Gaussian policy leads to some sudden drops in the reward function even in higher episodes and after training. This can be a harmful effect in our specific application scenario in factory automation where ensuring high-reliability is of paramount importance. Also, Fig. \ref{fig:rewards} shows the convergence of the TD3 method with deterministic policy in terms of difference number of RIS elements and BS transmit power. It is observed that, for either higher number of RIS elements or higher BS transmit power budget, the agent needs more episode to learn the optimized policy.

    
\begin{figure}[t]
     	\centering
     	\includegraphics[trim = 8.5cm 8.5cm 8.5cm 8.5cm,scale=0.6]{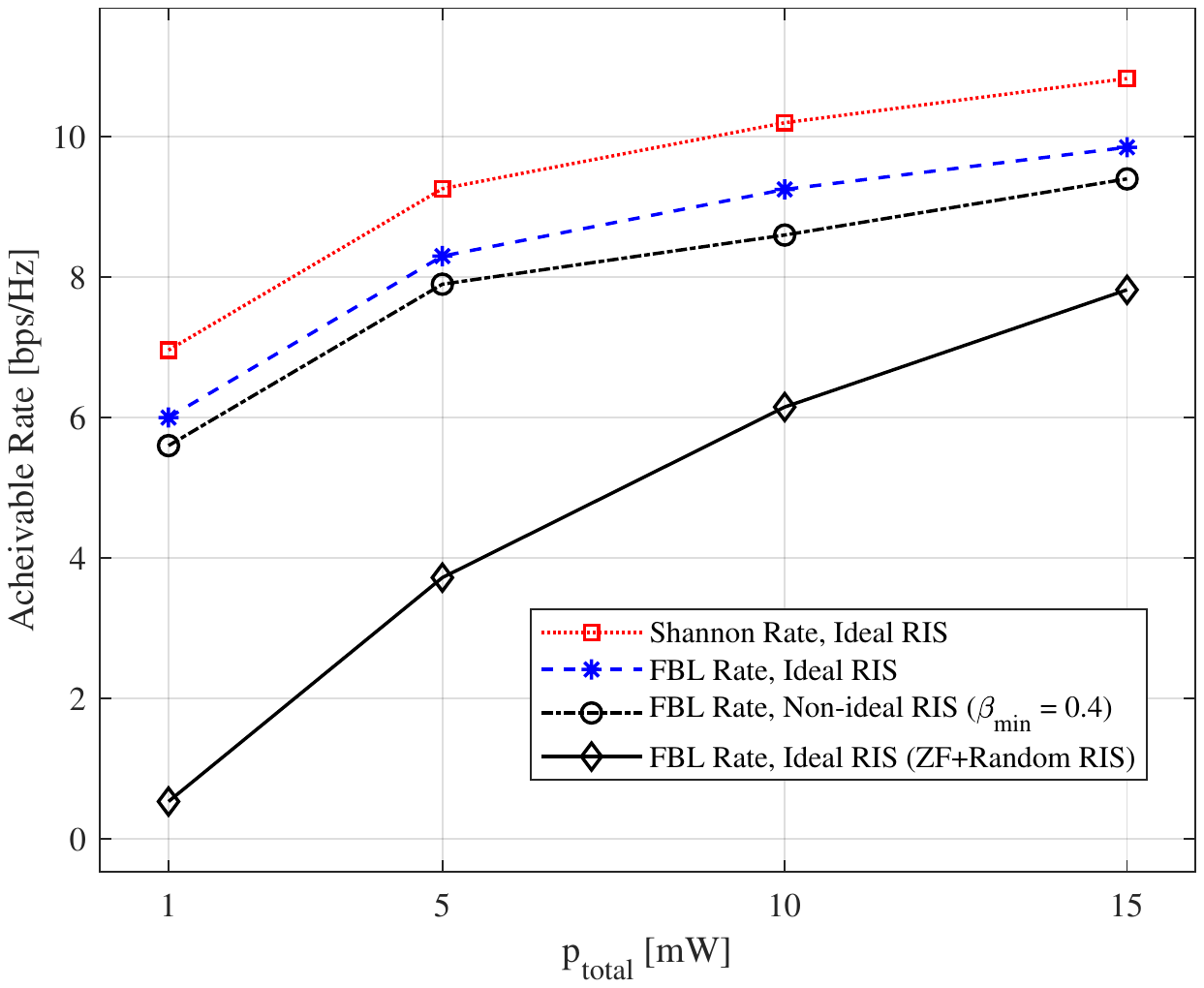}
       	\caption{The impact of increasing the BS transmit power on the converged average rate in FBL and Shannon regimes.}
       	\label{fig:FBL_vs_BS_transmit_power}
 	\end{figure}

	Fig. \ref{fig:FBL_vs_BS_transmit_power} shows the impact of increasing the BS transmit power on the average achievable rates in Shannon/FBL regime. As it is demonstrated, the uppermost red curve shows the case that the RIS is ideal and the Shannon capacity expression is leveraged illustrating the upperbound performance of the network in infinite CBL regime. It is also observed that increasing the transmit power budget at the BS leads a higher total rate in all scenarios. On the other side, the performance of the system in FBL regime with/without non-ideal RIS is illustrated in the lower curves. The achievable FBL rate by employing zero forcing (ZF) precoding at the BS and uniformly distributed random phase shift at the RIS is also shown in the lowermost curve for comparison. Note that the ZF performs better in higher SNR regimes as the gap between curves reduces, i.e., the ZF precoder and optimized CBL and active/passive beamformers curves get closer as the total transmit power $p_{\text{total}}$ increases. This highlights the applicability of our proposed resource allocation algorithm in system-level design considerations to establish reliable communications in industrial environments.
	
\begin{figure}[t]
 	    \centering
  		\includegraphics[trim = 8.5cm 8.5cm 8.5cm 8.5cm,scale=0.6]{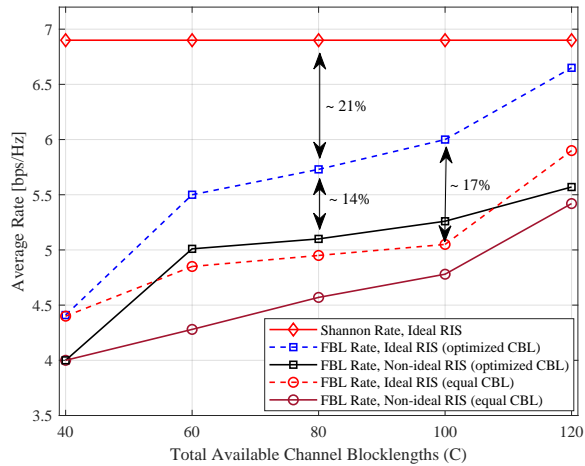}
  		\caption{The impact of increasing the total available CBL on the achivable FBL rate.}
 		\label{fig:FBL_CBL}
 	 \end{figure}

	Fig. \ref{fig:FBL_CBL}  shows the achievable rate performance comparison in terms of total available CBL. Since achievable rate expression in Shannon regime is independent from varying total CBL, the uppermost curve has no variations versus changing $C$. The performance gap between working in FBL regime and Shannon with either ideal or non-ideal RIS is also highlighted. There is a 21\% gap in ideal RIS case and 14\% extra penalty due to having non-ideal RIS. In addition, we have shown the case where the CBL variables are equally assigned between actuators, however the active and passive beamforming vectors are being optimized. It can be seen that there is around 17\% gap between CBL optimization and equal CBL allocation. From other perspective, the CBL can be expressed in terms of transmission duration $T$ and available bandwidth $W$ as $c=TW$. Thus, utilizing fewer CBLs results in decreasing the transmission duration. This shows the importance of optimizing the CBL to preserve the possible FBL rate loss and reduce the transmission time to meet URLLC KPIs. Note that when $C=40$ the optimized and non-optimzied curves overlap as the considered minimum CBL during simulations are $c_k^{\text{min}}=10$ $\forall k$, $\sum_{k=1}^{4}c_k^{\text{min}}=40$.
	
	
\begin{figure}[t]
 		\centering
 		\includegraphics[trim = 8.5cm 8.5cm 8.5cm 8.5cm,scale=0.6]{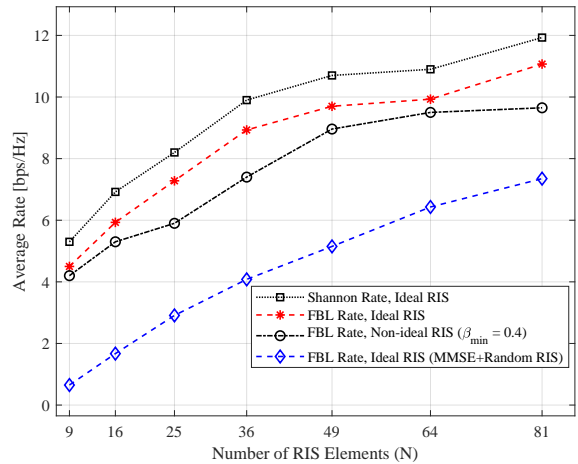}
 		\caption{The effect of increasing number of the RIS elements on the total achievable rate of the system.}
 		\label{fig:FBL_vs_N}
 	\end{figure}

	Similarly, in Fig. \ref{fig:FBL_vs_N} the network sum rate is assessed in terms of increasing the total number of reflective elements at the RIS. A gap is also observed between the Shannon achievable rate and FBL rate with either ideal or non-ideal RIS. The Shannon and FBL regime with non-ideal RIS curves demonstrate that the system actual performance will lie between these two curves. The performance of the proposed TD3 method is compared with state of the art linear minimum mean square error (MMSE) precoding at the BS. Moreover, the total achievable rate in all cases increases with the number of RIS elements, i.e., with/without ideal/non-ideal RIS. The similar performance is also shown in FBL and Shannon rates. On the other hand, the slope of the curves are quite similar when the number of RIS elements start to increase which additionally shows the practicality of the proposed TD3 algorithm in ideal/non-ideal reflective phase shift design problems. 

\begin{figure}[t]
 		\centering
 		\includegraphics[trim = 8.5cm 8.5cm 8.5cm 8.5cm,scale=0.6]{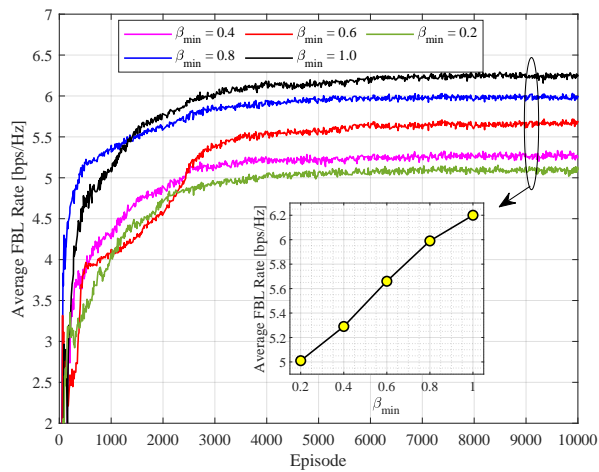}
 		\caption{The FBL rate in terms of $\beta_{\text{min}}$.}
 		\label{fig:FBL_vs_beta}
 	\end{figure}

 	Finally, Fig. \ref{fig:FBL_vs_beta} shows the effect of $\beta_{\text{min}}$ on the learning behaviour of the proposed TD3 agent. As can be seen from the curves, the agent reward function has converged in around 4000 episodes for all scenarios. In addition, there is a performance gap between $\beta_{\text{min}}=0.2$ and $\beta_{\text{min}}=1.0$ where the latter corresponds to the ideal RIS without amplitude attenuation. More precisely, the achievable FBL rate has increased from 5 bps/Hz ($\beta_{\text{min}}=0.2$) to 6.2 bps/Hz ($\beta_{\text{min}}=1$) which is a 20\% improvement.
 	

   \section{Conclusion}
   \label{ConslusionSec}
   We have studied the reflective phase shift design, BS beamforming and CBL allocation problem in practical RIS-aided URLLC systems over short packet communications. The RIS impairments are modeled as the non-linear amplitude response via changing the phase shift values, and the considered problem has been addressed by a novel and efficient DRL algorithm. First, we formulated the optimization problem framework with the objective of maximizing total FBL rate while meeting a given target reliability of actuators in a factory environment proposed where the constraints are the elements' amplitude response in terms of phase shift values, total available CBL for actuators and BS transmit power budget. Since the proposed problem has highly non-linear constraints due to considering practical phase shift response, it is relatively challenging to solve via optimization-based algorithms that are usually computationally inefficient even in ideal scenarios. Thus, we have employed a policy gradient DRL algorithm based on unsupervised actor-critic methods to optimize the active/passive beamforming and CBL allocation which concurrently learns a Q-function and a policy. The utilized DRL method, i.e., TD3 has addressed the issues in the conventional DDPG method that overestimate action-value function, which then leads to the policy breaking. The numerical results have demonstrated the applicability of the proposed DRL method in practical RIS phase shift design problems in time-sensitive applications that exploit short packets in URLLC systems. Moreover, the proposed TD3 method with deterministic policy outperformed other considered DRL algorithms such as soft actor-critic and Gaussian policy randomization in terms of final reward values and generalization of the policy network for different channel coefficients. In addition, we investigated the importance of optimizing the CBL in short packet communications and showed that the system total FBL can increase by 17\% when the CBL variables are optimized for each actuator.

    
    
	\bibliographystyle{IEEEtran}
    \bibliography{IEEEabrv,refs}

	
\end{document}